\documentclass[10pt, conference]{IEEEtran}

\IEEEoverridecommandlockouts
\usepackage{amsmath,amssymb,amsfonts}
\usepackage{algorithmic}
\usepackage{graphicx}
\usepackage{textcomp}
\usepackage{xcolor}
\usepackage{color, soul}
\usepackage{hyperref}
\usepackage{booktabs}
\usepackage{multirow}
\usepackage{url}
\usepackage[style=ieee, url=true]{biblatex}

\AtEveryBibitem{%
  \clearfield{language}%
  \clearlist{language}%
  
  \clearfield{doi}%
  \clearfield{note}%
  \clearfield{issn}%
  \clearfield{month}%
  \clearfield{pagetotal}%
  \clearname{translator}%
  
  \ifentrytype{online}{}{%
    \ifentrytype{misc}{}{%
      \ifentrytype{software}{}{%
        \clearfield{url}%
        \clearfield{urlyear}%
      }%
    }%
  }%
}

\hypersetup{
  colorlinks=true,
  linkcolor=blue,
  citecolor=blue,
  urlcolor=blue
}

\usepackage{tikz}

\usetikzlibrary{arrows.meta,positioning,shapes.geometric,fit,backgrounds,decorations.pathreplacing,calc}

\begin{document}

\title{\underline{Q}uantum \underline{A}lgorithm for \underline{D}istributed \underline{R}eduction of Entanglements (QADR): A Trainable and Simulation-Efficient QML Framework}

\author{
\IEEEauthorblockN{Syed Farhan Ahmad\IEEEauthorrefmark{1}, Gregory T. Byrd\IEEEauthorrefmark{1}}
\IEEEauthorblockA{\IEEEauthorrefmark{1}\textit{Department of ECE, North Carolina State University, Raleigh, North Carolina, USA} \\
\{sahmad25, gbyrd\}@ncsu.edu
}
}

\maketitle

\begin{abstract}
Training Variational Quantum Circuits (VQCs) under Noisy Intermediate-Scale Quantum (NISQ) constraints introduces severe computational limitations: classical statevector simulation memory scales exponentially ($\mathcal{O}(2^n)$), and global cost functions suffer from barren plateaus where gradient variance decays exponentially ($\mathcal{O}(1/2^n)$). This paper introduces and evaluates the Quantum Algorithm for Distributed Reduction of Entanglements (QADR), a hybrid quantum-classical machine learning framework that decomposes a global $n$-qubit VQC into localized sub-circuits operating approximately within the causal light cones of individual target qubits. QADR reduces classical simulation memory scaling from $\mathcal{O}(2^n)$ to $\mathcal{O}(n \cdot 2^{2d+1})$ for a light cone radius $d$, while naturally mitigating global barren plateaus. We benchmark QADR against standard global VQCs, Support Vector Machines (SVM), and two customized classical parameter-matched neural networks (CANN and PMNN) on the MNIST dataset and the high-dimensional NASA IMS wind turbine drivetrain diagnostic task. QADR demonstrates excellent scalability, operating successfully at $n_{\text{features}}=2000$ where standard global VQCs crash due to memory exhaustion, while matching or exceeding the performance of optimized classical architectures.
\end{abstract}


\section{Introduction}

Supervised Quantum Machine Learning (QML) utilizes parameterized quantum unitaries within highly expressible Hilbert spaces to identify complex classification boundaries \cite{rebentrost_quantum_2014, havlicek_supervised_2019}. Despite strong theoretical support, classical high-performance computing (HPC) statevector simulators encounter a rigid memory limit: storing and tracking a full statevector across $n$ qubits requires $2^n$ complex amplitudes, rendering unconstrained systems exceeding roughly 30 to 36 qubits classically unsimulable.

Beyond classical simulation limits, optimizing deep VQCs is fundamentally hard: global cost functions suffer an exponential decay in gradient variance, $\operatorname{Var}[\partial \mathcal{L}/\partial \theta_k] \propto 1/2^n$, the barren plateau phenomenon \cite{mcclean_barren_2018}. Shifting to local cost functions restricts gradient degradation to a polynomial boundary at shallow depths \cite{cerezo_variational_2021}, yet global optimization landscapes remain difficult to navigate when all parameters are updated concurrently over a fully entangled statevector.

This work addresses these joint bottlenecks by developing and evaluating the QADR. By using approximate causal light cones to restrict entangling structures around target qubits, QADR replaces a monolithic wide VQC with an ensemble of localized, low-width, fixed-overhead quantum tasks supervised by a lightweight classical orchestrator. To ensure a mathematically rigorous evaluation of QADR's performance, we compare its performance against two custom classical neural baselines designed as direct parameter-matched ablations of the quantum feature map.

\section{QADR Architectural Design}

\subsection{Causal Light Cones in Hardware-Efficient Layouts}
In a standard 1D linear nearest-neighbor qubit topology evolved under an alternating-layer hardware-efficient ansatz, evaluating a strictly localized single-qubit observable $O_i = Z_i$ on a target qubit $i$ ensures that gates executing outside the backward causal path leave the expectation value invariant due to unitary cancellation ($U^\dagger U = I$)~\cite{huang_learning_2024}.

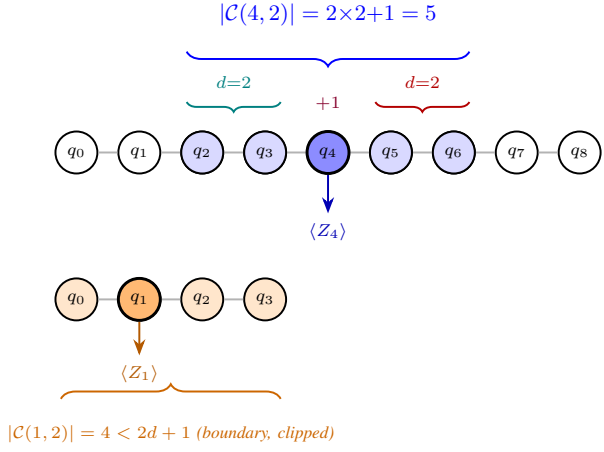
\begin{figure}[!t]
\centering
\resizebox{0.9\columnwidth}{!}{%
\begin{tikzpicture}[>=Stealth, font=\scriptsize, node distance=0pt]
  \def\sp{0.9}    
  \def\nd{0.60cm} 

  \foreach \i in {0,...,8} {
    \node[draw, circle, thick, fill=white,
          minimum size=\nd, inner sep=0pt]
      (q\i) at (\i*\sp, 0) {$q_{\i}$};
  }
  \foreach \i [evaluate=\i as \j using int(\i+1)] in {0,...,7} {
    \draw[thick, gray!60] (q\i) -- (q\j);
  }

  \foreach \i in {2,3,5,6} {
    \node[draw, circle, thick, fill=blue!15,
          minimum size=\nd, inner sep=0pt]
      (q\i) at (\i*\sp, 0) {$q_{\i}$};
  }
  \node[draw, circle, very thick, fill=blue!45,
        minimum size=\nd, inner sep=0pt]
    (q4) at (4*\sp, 0) {$q_4$};

  \draw[decorate, decoration={brace, amplitude=4pt, mirror},
        thick, teal]
    ($(q2.north west)+(0,0.50)$) -- ($(q3.north east)+(0,0.50)$)
    node[midway, above=2pt, teal] {$d{=}2$};

  \node[above=0.16cm, purple!70!black] at (q4.north) {$+1$};

  \draw[decorate, decoration={brace, amplitude=4pt, mirror},
        thick, red!70!black]
    ($(q5.north west)+(0,0.5)$) -- ($(q6.north east)+(0,0.5)$)
    node[midway, above=2pt, red!70!black] {$d{=}2$};

  \draw[decorate, decoration={brace, amplitude=6pt, mirror},
        thick, blue]
    ($(q2.north west)+(0,1.22)$) -- ($(q6.north east)+(0,1.22)$)
    node[midway, above=7pt, blue, font=\small]
      {$|\mathcal{C}(4,2)| = 2{\times}2{+}1 = 5$};

  \draw[->, thick, blue!70!black]
    (q4.south) -- +(0,-0.55)
    node[below] {$\langle Z_4\rangle$};

  \foreach \i in {0,...,3} {
    \node[draw, circle, thick, fill=orange!20,
          minimum size=\nd, inner sep=0pt]
      (bq\i) at (\i*\sp, -2.1) {$q_{\i}$};
  }
  \foreach \i [evaluate=\i as \j using int(\i+1)] in {0,...,2} {
    \draw[thick, gray!60] (bq\i) -- (bq\j);
  }
  \node[draw, circle, very thick, fill=orange!55,
        minimum size=\nd, inner sep=0pt]
    (bq1t) at (1*\sp, -2.1) {$q_1$};

  \draw[decorate, decoration={brace, amplitude=6pt},
        thick, orange!80!black]
    ($(bq0.south west)+(0,-1.20)$) -- ($(bq3.south east)+(0,-1.20)$)
    node[midway, below=7pt, orange!80!black]
      {$|\mathcal{C}(1,2)| = 4 < 2d+1$ \textit{(boundary, clipped)}};


  \draw[->, thick, orange!70!black]
    (bq1t.south) -- +(0,-0.48)
    node[below] {$\langle Z_1\rangle$};

\end{tikzpicture}%
}
\caption{Causal cone geometry for $n=9$, $d=2$.
  \textbf{Top (interior cone):} target $q_4$ sits at the center of a
  full-width $2d{+}1 = 5$-qubit cone ($d$ qubits left + target + $d$ right).
  \textbf{Bottom (boundary cone):} target $q_1$ is only one hop from the left
  edge, so the cone is clipped to four qubits, with one left neighbor instead of two.
  There are always exactly $2d$ boundary cones.}
\label{fig:cone_geometry}
\end{figure}

For a hardware-efficient ansatz configuration operating at an interaction layer depth $d$, the causal light cone $\mathcal{C}(i, d)$ containing the set of indices that can mathematically affect the expectation value of target qubit $i$ is formulated as:
\begin{equation}
\mathcal{C}(i, d) = \{ j \in \mathbb{Z} \mid \max(0, i-d) \leq j \leq \min(n-1, i+d) \}
\end{equation}
Because the entangling layers expand symmetrically to both the left and right sides of the target node, the maximal width of an interior sub-register is bounded by:
\begin{equation}
|\mathcal{C}(i,d)| = \underbrace{d}_{\text{left}} + \underbrace{1}_{\text{center}} + \underbrace{d}_{\text{right}} = 2d+1
\end{equation}
At the physical boundaries of the chain ($i < d$ or $i > n-1-d$), the sub-registers are naturally clipped by the hardware perimeter, creating exactly $2d$ boundary cones of smaller width. This structural containment is depicted in Figure \ref{fig:cone_geometry}. 

\subsection{Distributed Architecture and the Classical Orchestrator}
Instead of instantiating an $n$-qubit global circuit, QADR dynamically generates $n$ distinct, localized sub-circuits. Each sub-circuit $i$ maps only the inputs and parameter subsets relevant to $\mathcal{C}(i, d)$. 

\begin{figure*}[t]
\centering
\begin{tikzpicture}[
  block/.style={draw, rectangle, fill=blue!10, thick, minimum width=2.4cm, minimum height=0.7cm, align=center},
  orch_block/.style={draw, rectangle, fill=green!10, thick, minimum width=2.2cm, minimum height=3.8cm, align=center},
  input/.style={circle, draw, fill=gray!10, minimum size=1.1cm, align=center},
  expect/.style={circle, draw, fill=red!10, minimum size=1.1cm, align=center},
  output/.style={circle, draw, fill=yellow!20, minimum size=1.1cm, align=center},
  arrow/.style={-Latex, thick}
]
  \node[input] (x1) at (-3.5, 3.0) {$x_0$};
  \node[input] (x2) at (-3.5, 1.6) {$x_1$};
  \node at (-3.5, 0.6) {$\vdots$};
  \node[input] (xn) at (-3.5, -0.4) {$x_{n-1}$};
  \draw[thick, red!40, dashed, rounded corners] (-4.3, 3.6) rectangle (-2.7, -1.0);
  \node[red!80!black, font=\bfseries] at (-3.5, 3.9) {Inputs $\mathbf{x}$};

  \draw[arrow, line width=1.5pt, blue!70!black] (-2.4, 1.3) -- (-0.6, 1.3) node[midway, above, font=\bfseries] {QADR};

  \node[block] (sc1) at (1.5, 3.0) {Sub-circuit $0$ \\ $\mathcal{C}(0, d)$};
  \node[block] (sc2) at (1.5, 1.6) {Sub-circuit $1$ \\ $\mathcal{C}(1, d)$};
  \node at (1.5, 0.6) {$\vdots$};
  \node[block] (scn) at (1.5, -0.4) {Sub-circuit $n{-}1$ \\ $\mathcal{C}(n{-}1, d)$};
  \draw[thick, blue!30, dashed, rounded corners] (0.1, 3.6) rectangle (2.9, -1.0);
  \node[blue!80!black, font=\bfseries] at (1.5, 3.9) {Local Unitaries};

  \node[expect] (e1) at (4.5, 3.0) {$E_0$};
  \node[expect] (e2) at (4.5, 1.6) {$E_1$};
  \node at (4.5, 0.6) {$\vdots$};
  \node[expect] (en) at (4.5, -0.4) {$E_{n-1}$};

  \draw[arrow] (sc1) -- (e1);
  \draw[arrow] (sc2) -- (e2);
  \draw[arrow] (scn) -- (en);
  
  \draw[thick, red!30, dashed, rounded corners] (3.7, 3.6) rectangle (5.3, -1.0);
  \node[red!80!black, font=\bfseries] at (4.5, 3.9) {Expectations};

  \node[orch_block] (orch) at (7.2, 1.3) {Classical\\Orchestrator};
  \draw[thick, green!40, dashed, rounded corners] (5.8, 3.6) rectangle (8.6, -1.0);
  \node[green!60!black, font=\bfseries] at (7.2, 3.9) {Orchestration};

  \draw[arrow] (e1) -- (orch.west |- e1);
  \draw[arrow] (e2) -- (orch.west |- e2);
  \draw[arrow] (en) -- (orch.west |- en);

  \node[output] (yhat) at (9.6, 1.3) {$\hat{y}$};
  \draw[arrow] (orch.east) -- (yhat);
  
\end{tikzpicture}
\caption{QADR Pipeline: an $n$-qubit classification task is decomposed into $n$ fixed-width local sub-circuits, each operating within its causal light cone $\mathcal{C}(i,d)$. Local $\langle Z_{target} \rangle$ expectation values are fused by a lightweight Classical Orchestrator, reducing simulation memory from $\mathcal{O}(2^n)$ to $\mathcal{O}(n \cdot 2^{2d+1})$ while preserving end-to-end trainability.}
\label{fig:qadr_flow}
\end{figure*}
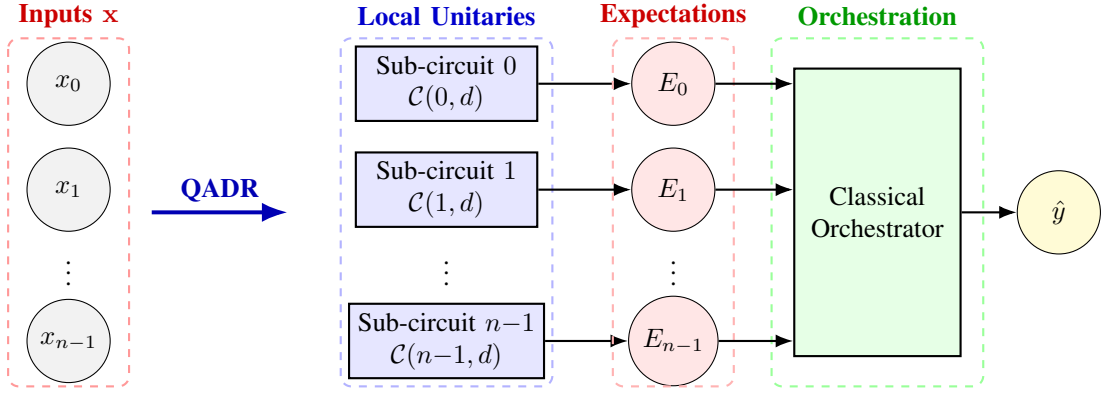

The individual local expectations are evaluated across all sub-registers and gathered into a structural intermediate feature array $\mathbf{e}$:
\begin{equation}
\mathbf{e} = [E_0, E_1, \dots, E_{n-1}]^T, \quad E_i = \langle \psi(\theta_i) | Z_{target} | \psi(\theta_i) \rangle
\end{equation}
This representation is passed to a lightweight Classical Orchestrator consisting of a dense projection layer with parameterized weights $\mathbf{W}_1 \in \mathbb{R}^{8 \times n}$, a bias $\mathbf{b}_1 \in \mathbb{R}^8$, and a binary classification sigmoid layer with weights $\mathbf{w}_2 \in \mathbb{R}^8$ and bias $b_2$:
\begin{equation}
\hat{y} = \sigma\!\bigl(\mathbf{w}_2^{\top}\,\mathrm{ReLU}(\mathbf{W}_1\mathbf{e} + \mathbf{b}_1) + b_2\bigr)
\end{equation}
The global model parameters are trained end-to-end via gradient-based backpropagation using the standard binary cross-entropy loss, as depicted in Figure~\ref{fig:qadr_flow}.

\subsection{Circuit Ansatz and Approximate Causal Light Cones}
All VQC models share a hardware-efficient ansatz~\cite{kandala_hardware-efficient_2017}:
\begin{enumerate}
  \item \textbf{Encoding:} $R_x(x_i)$ for each qubit $i$
  \item \textbf{Variational layers} ($L=2$, repeated):
    \begin{itemize}
      \item Linear nearest-neighbor CNOT entanglement: $\text{CNOT}(i, i+1)$
      \item Trainable rotations: $R_x(\theta_{i,0})\,R_y(\theta_{i,1})\,R_z(\theta_{i,2})$
    \end{itemize}
\end{enumerate}

Each circuit applies the
shared ansatz to its cone qubits, measuring $\langle Z_{\text{target}}\rangle$
at the cone's center (and on qubits 2d around the center).

\subsection{Complexity Analysis and the Classical Simulation Memory Wall}
The computational performance benefits of the QADR layout are derived through statevector storage constraints.
\\
\noindent\textbf{Theorem 1 (Simulation Resource Scaling):} \textit{Let $n$ be the total input feature count mapped to $n$ qubits, and $d$ be the chosen light cone layer radius. The classical statevector allocation bounds translate to:}
\begin{equation}
\text{Memory}_{\text{Global}} = \mathcal{O}(2^n), \quad \text{Memory}_{\text{QADR}} = \mathcal{O}(n \cdot 2^{2d+1})
\end{equation}
\textit{Proof:} A standard global statevector tracks a full state space tracking $2^n$ complex coefficients. QADR isolates execution to $n$ individual allocations, where each allocation contains at most $2d+1$ active qubits. The total structural amplitude tracking space scales as $n \times 2^{2d+1}$. Because $d \ll n$ remains a fixed structural parameter, storage scales linearly ($\mathcal{O}(n)$) rather than exponentially ($\mathcal{O}(2^n)$).

For example, compiling an operational model at $n=50$ features with a cone depth of $d=2$ would require a classical global allocation of $2^{50} \approx 1.12 \times 10^{15}$ amplitudes, demanding over 16 petabytes of RAM. QADR compresses this same target framework into $50 \times 2^5 = 1,600$ amplitudes, requiring less than 2 KB of memory.

\section{Controlled Classical Ablations}
To isolate the contribution of the quantum feature map, we benchmark QADR against two customized classical neural architectures that share the exact same orchestrator head as QADR.

\subsection{Classical Analogous Neural Network (CANN)}
CANN replaces the quantum feature block with a single trainable affine layer of the same output dimensionality, leaving the classical head unchanged:
\begin{equation}
  \mathbf{f}(\mathbf{x};\, W, \mathbf{b}) = \mathrm{GELU}\!\bigl(\mathrm{BN}(W\mathbf{x})\bigr), \quad W \in \mathbb{R}^{n \times n}
\end{equation}
where $\mathrm{BN}(\cdot)$ denotes pre-activation Batch Normalization and $\mathrm{GELU}(\cdot)$ represents the Gaussian Error Linear Unit activation. The $n \times n$ weight matrix $W$ is structurally analogous to the $n$-qubit global VQC, mapping an $n$-dimensional input vector to an $n$-dimensional feature vector. This feature block contributes $n^2 + 2n$ parameters (representing $\mathcal{O}(n^2)$ scaling).

\subsection{Parameter Matched Neural Network (PMNN)}
The $\mathcal{O}(n^2)$ feature block of CANN becomes an unfair competitor against QADR at large $n$ because QADR's parameter count scales linearly ($\mathcal{O}(n)$). To establish a mathematically fair baseline, PMNN matches QADR's parameter count exactly via a two-layer encoder structure:
\begin{align}
  \mathbf{h}(\mathbf{x}) &= \mathrm{GELU}\!\bigl(\mathrm{BN}(W^{(1)}\mathbf{x})\bigr), \quad W^{(1)} \in \mathbb{R}^{h \times n}\\
  \mathbf{f}(\mathbf{x}) &= \mathrm{GELU}(W^{(2)}\mathbf{h} + \mathbf{b}^{(2)}), \quad W^{(2)} \in \mathbb{R}^{n \times h}
\end{align}
followed by the same orchestrator head. The encoder maps input $n \to h \text{ (hidden)} \to n$, recovering an $n$-dimensional feature vector. The feature-block parameter count is:
\begin{equation}
  P_{\text{C2}} = h(2n+2) + n
\end{equation}
where $2h$ accounts for the BN scale and shift of the hidden layer. To determine the hidden width $h$ for a given $n$, we set $P_{\text{C2}} = P_{\text{QADR}}$ and solve for $h$:
\begin{equation}
  h = \left\lfloor \frac{P_{\text{QADR}} - n}{2n+2} \right\rfloor
\end{equation}
This ensures that the classical model possesses no parameter advantage over QADR during comparative evaluation. The total model parameter counts across scaling input sizes are listed in Table \ref{tab:param-counts}.

\begin{table}[h]
\centering
\caption{Model Parameter Count Comparison}
\label{tab:param-counts}
\begin{tabular}{rcccc}
\toprule
$n$ & \textbf{Glob. VQC} & \textbf{QADR} & \textbf{CANN} & \textbf{PMNN} \\
\midrule
 5 &  87  &  171 &   92 &  170 \\
10 & 157  &  361 &  217 &  371 \\
15 & 227  &  551 &  392 &  536 \\
20 & 297  &  741 &  617 &  743 \\
50 & 717  & 1881 & 3017 & 1895 \\
\bottomrule
\end{tabular}
\end{table}

\section{Experimental Methodology \& Datasets}

All models were evaluated using a 5-fold cross-validation scheme over 3 independent random shuffles (15 total fits per configuration) to minimize dependency on fold assignments. 

\subsection{MNIST Sanity Check}
To confirm theoretical convergence, a baseline study was conducted on a binary subset of the MNIST handwritten digit dataset (digits 0 and 1). Flat images were downsampled to $n_q=10$ using Principal Component Analysis (PCA).

\subsection{NASA IMS Drivetrain Diagnostics}
To test the high-dimensional scaling limits of QADR against classical architectures, we use the NASA Inductive Monitoring System (IMS) bearing vibration dataset~\cite{qiu_wavelet_2006}, specifically the \texttt{2nd\_test} run monitoring Bearing~1, which develops an outer race fault. The dataset provides 400 balanced samples (200 normal, 200 incipient-fault), each consisting of a 2,048-point vibration measurement window. A stratified 75/25 train-test split (300 train, 100 test) is applied before cross-validation. All reported metrics are averaged over a Repeated Stratified 5-Fold scheme with 3 independent random shuffles (15 total fits per configuration). For $n \leq 400$, PCA is applied; for $n > 400$ (e.g.\ $n=1000, 2000$), sklearn's \texttt{SelectKBest} is used instead, since fitting a PCA basis larger than the sample count is ill-posed. All feature vectors are rescaled to $[0, \pi]$ using a per-fold \texttt{MinMaxScaler}.

\section{Experimental Results}

\subsection{Simulation Efficiency Benchmarks}
We compared the wall-clock execution times of QADR training in both serial and parallel configurations. 

\begin{figure}[t]
\centering
\centering
\includegraphics[width=0.8\columnwidth]{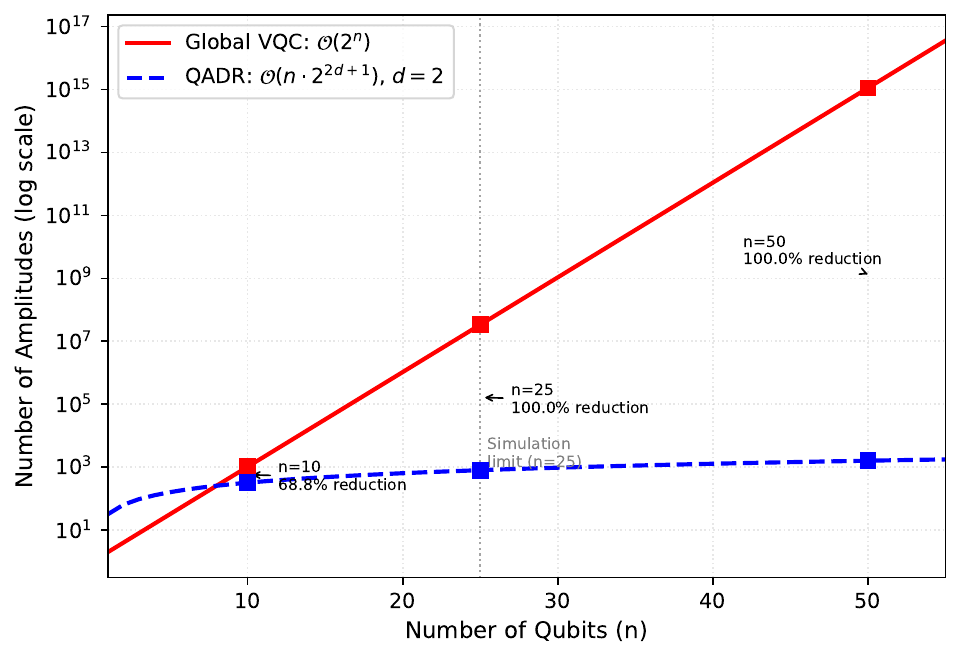}
\caption{Theoretical statevector simulation amplitude complexity comparison between Global VQC ($\mathcal{O}(2^n)$) and QADR ($\mathcal{O}(n \cdot 2^{2d+1})$) with a cone radius of $d=2$ on a logarithmic scale.}
\label{fig:scaling_theory_tikz}
\end{figure}

The serial configuration loops through each of the $n$ sub-registers sequentially, whereas the parallel configuration utilizes batched tensor executions via TensorFlow \texttt{vmap}-style graph fusion across all cones simultaneously. As shown in Table~\ref{tab:speedup-results}, the parallelized QADR execution offers significant speedups, growing from $1.1\times$ at $n=5$ to $13.6\times$ at $n=50$ and reaching $51.3\times$ at $n=200$, while maintaining peak RAM consumption well under $4$~GB regardless of $n$.

\begin{table}[h]
\centering
\caption{QADR Training Execution Time (20 Epochs, $d=2$)}
\label{tab:speedup-results}
\begin{tabular}{cccc}
\toprule
$n$ & \textbf{Serial Time (s)} & \textbf{Parallel Time (s)} & \textbf{Speedup Factor} \\
\midrule
  5 &    76.3 &  70.0 &  $1.1\times$ \\
 10 &   173.5 &  74.0 &  $2.3\times$ \\
 20 &   415.2 &  98.0 &  $4.2\times$ \\
 50 &  1136.7 &  83.7 & $13.6\times$ \\
100 &  2316.3 &  84.4 & $27.5\times$ \\
200 &  5024.0 &  97.8 & $51.3\times$ \\
\bottomrule
\end{tabular}
\end{table}

\subsection{Classification Performance}
On the binary MNIST task ($n_q=10$), all architectures successfully converged, with QADR securing a classification accuracy of $100\%$ and the Global VQC achieving $99.50\%$.

\begin{table}[h]
\centering
\caption{NASA IMS Fault Classification Generalization Metrics}
\label{tab:ims-results}
\begin{tabular}{lcccc}
\toprule
\textbf{Model} & $n_{\text{features}}$ & \textbf{Accuracy (\%)} & \textbf{F1-Score} & \textbf{ROC-AUC} \\
\midrule
SVM (RBF) & 15 & 96.00 & 0.9602 & 0.9927 \\
CANN & 15 & 72.17 & 0.7761 & 0.9032 \\
PMNN & 15 & 74.25 & 0.7873 & 0.8879 \\
Global VQC & 15 & 83.83 & 0.8418 & 0.9275 \\
QADR ($d=2$) & 15 & 92.50 & 0.9222 & 0.9695 \\
\midrule
SVM (RBF) & 2000 & 96.08 & 0.9623 & 0.9958 \\
QADR ($d=2$) & 2000 & \textbf{99.25} & \textbf{0.9925} & \textbf{0.9995} \\
\bottomrule
\end{tabular}
\end{table}

The classification performance on the NASA IMS task is detailed in Table \ref{tab:ims-results}. At a feature-space dimension of  $n_{\text{features}}=15$, QADR achieved an accuracy of $92.50\%$ and an ROC-AUC of $0.9695$, outperforming both the standard Global VQC ($83.83\%$) and classical parameter-matched neural networks (CANN: $72.17\%$, PMNN: $74.25\%$), while SVM achieved the top accuracy of $96.00\%$ in this low-dimensional regime.

When evaluated at the high-dimensional scaling target of $n_{\text{features}}=2000$ (SelectKBest), standard Global VQCs and parameter-matched neural networks were excluded due to memory constraints and inability to classify in a regime where a ``curse of dimensionality" exists respectively. QADR, however, scaled seamlessly within its linear complexity bounds, attaining a classification accuracy of $99.25\%$ and an ROC-AUC of $0.9995$. This significantly outpaces the optimized SVM ($96.08\%$). These results demonstrate that the localized spatial causality enforced by QADR's light cones provides an exceptionally effective inductive bias for high-dimensional classification tasks.

\begin{figure}[h]
\centering
\includegraphics[width=0.75\columnwidth]{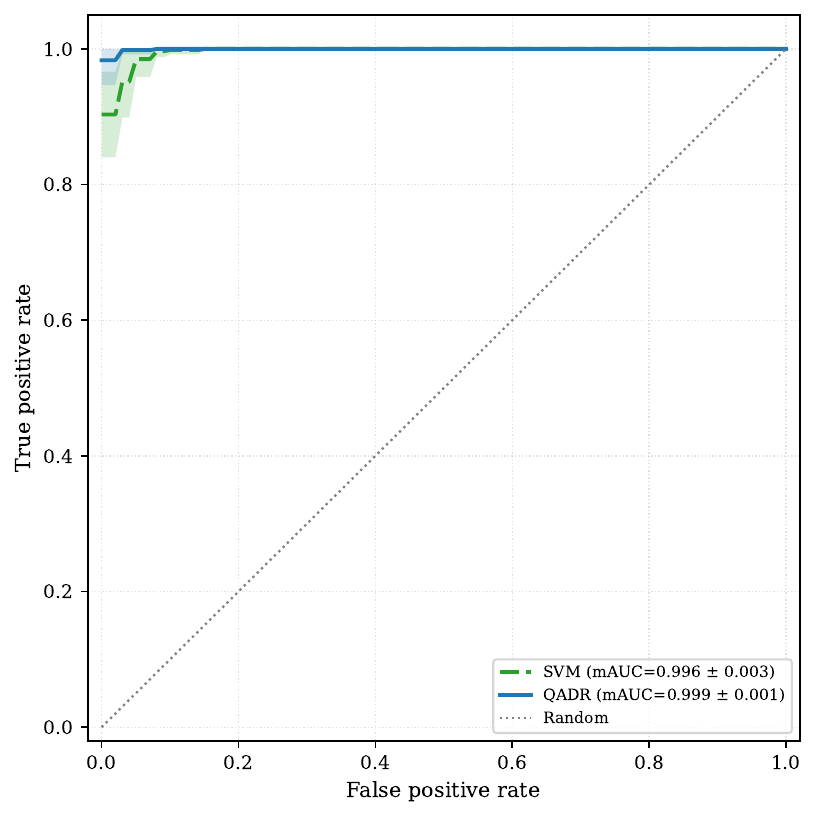}
\caption{ROC curves for QADR and SVM at $n_{\text{features}}=2000$ (SelectKBest) on the NASA IMS fault classification task. Only these two architectures were capable of operating at this scale where a ``curve of dimensionality" exists. QADR achieves an ROC-AUC of $0.9995$ versus $0.9958$ for SVM.}
\label{fig:roc_ims_2000}
\end{figure}

\section{Conclusion}
This paper introduces and validates QADR as a scalable and trainable framework for high-dimensional Quantum Machine Learning. By decomposing a monolithic circuit into localized sub-registers using approximate causal light cones, QADR bypasses the classical simulation memory wall while mitigating barren plateaus. Benchmarks against custom parameter-matched classical architectures confirm that QADR's performance gains are driven by a structurally robust quantum inductive bias, establishing a viable pathway for large-scale near-term quantum utility.

\section*{Acknowledgement}

During the preparation of this thesis, the authors utilized interactive generative AI assistants, specifically Google Gemini 3.1\footnote{\url{https://gemini.google.com}}, Anthropic Claude 4.6 Sonnet\footnote{\url{https://www.anthropic.com/claude}}, and the Cursor Composer suite (versions 2.0 and 2.5)\footnote{\url{https://www.cursor.com}}, to support various structural workflows. All technical content, mathematical derivations, citations, and conclusions were verified by and are the full responsibility of the authors.

\vspace{12pt}

\printbibliography

\end{document}